\documentclass[aps,prb,twocolumn,showpacs,showkeys,amsmath,amssymb,superscriptaddress]{revtex4-1}
\usepackage[T1]{fontenc}
\usepackage[latin9]{inputenc}
\usepackage{graphicx}
\usepackage{subfig}

\usepackage{dcolumn}% Align table columns on decimal point
\usepackage{bm}% bold math

\usepackage{epsfig}
\usepackage{subfig}
\usepackage{natbib}

\usepackage{bm}
\usepackage{multirow}
\usepackage{xcolor}
\usepackage{tabu}

\begin{document}

%\articletype{ARTICLE TEMPLATE}% Specify the article type or omit as appropriate

\title{Relative relevance of mobility and driving force on edge dislocation climb by the vacancy mechanism}

\author{Enrique Mart\'inez}
\email{enrique@clemson.edu}
\affiliation{Department of Mechanical Engineering, Clemson University, Clemson, SC 29623 USA}
\affiliation{Department of Materials Science and Engineering, Clemson University, Clemson, SC 29623 USA}

\author{Alankar Alankar}
\affiliation{Department of Mechanical Engineering, Indian Institute of Technology, Bombay, Mumbai, Maharashtra 400076, India}
%
%\author{John P. Hirth}
%
%\affiliation{Material Science and Technology Division, MST-8, Los Alamos National Laboratory, Los Alamos, 87545 NM, USA}
%
%\author{Arthur F. Voter}
%
%
%\affiliation{Theoretical Division, T-1, Los Alamos National Laboratory, Los Alamos, 87545 NM, USA}
%
\author{Alfredo Caro}
\affiliation{College of Professional Studies, George Washington University, Ashburn, Virginia 20147, USA}
%\affiliation{Los Alamos National Laboratory, Los Alamos, 87545 NM, USA}
\author{Thomas Jourdan}
\email{thomas.jourdan@cea.fr}
\affiliation{Universit\'e Paris-Saclay, CEA, Service de Recherches de M\'etallurgie Physique, 91191, Gif-sur-Yvette, France}

\date{\today}

%\date{\today}
\begin{abstract}
In this work we examine the driving force for edge dislocations to climb in $\alpha$-Fe from atomistic and mesoscale viewpoints. We study the bias for the climb process depending on the dislocation orientation and the applied stress as coming from both the gradient of the chemical potential and the transport coefficients. Both terms are modified by the applied stress and therefore contribute to climb. Surprisingly, even though the vacancy migration barrier distribution is modified by the external stress as obtained by nudged-elastic band calculations, the mobilities resulting from a kinetic Monte Carlo model applied on the obtained energy landscape are indistinguishable, independently of the stress. Moreover, an object kinetic Monte Carlo (OKMC) model including the effect of the dislocation strain field to first order shows indeed a slight anisotropic component in the diffusion in more complex dislocation configurations. However, the OKMC results highlight the fact that the thermodynamic component is the dominant driving force. We conclude that in $\alpha$-Fe under thermal conditions, the main source of bias is given by the difference in vacancy chemical potentials, which is small enough to hinder the process for dynamic atomistic simulations. %[\textcolor{blue}{Do you mean that with atomistic simulations, you cannot see the effect of the chemical potential? Is this a result that you obtain from the KMC simulations? In my view the KMC results tell you more about the kinetic effect (stress-induced change of the diffusion), right? Otherwise you should describe atomistically the emission of point defects by dislocations}].

\end{abstract}

\maketitle

%\pacs{\textbf{68.35.-p,68.35.Dv,61.72.J-,61.80.Az}}
%\keywords{\textbf{Dislocations, Climb, Mechanical Response}}
%
%\maketitle

\begin{figure}
\includegraphics[width=1.0\columnwidth]{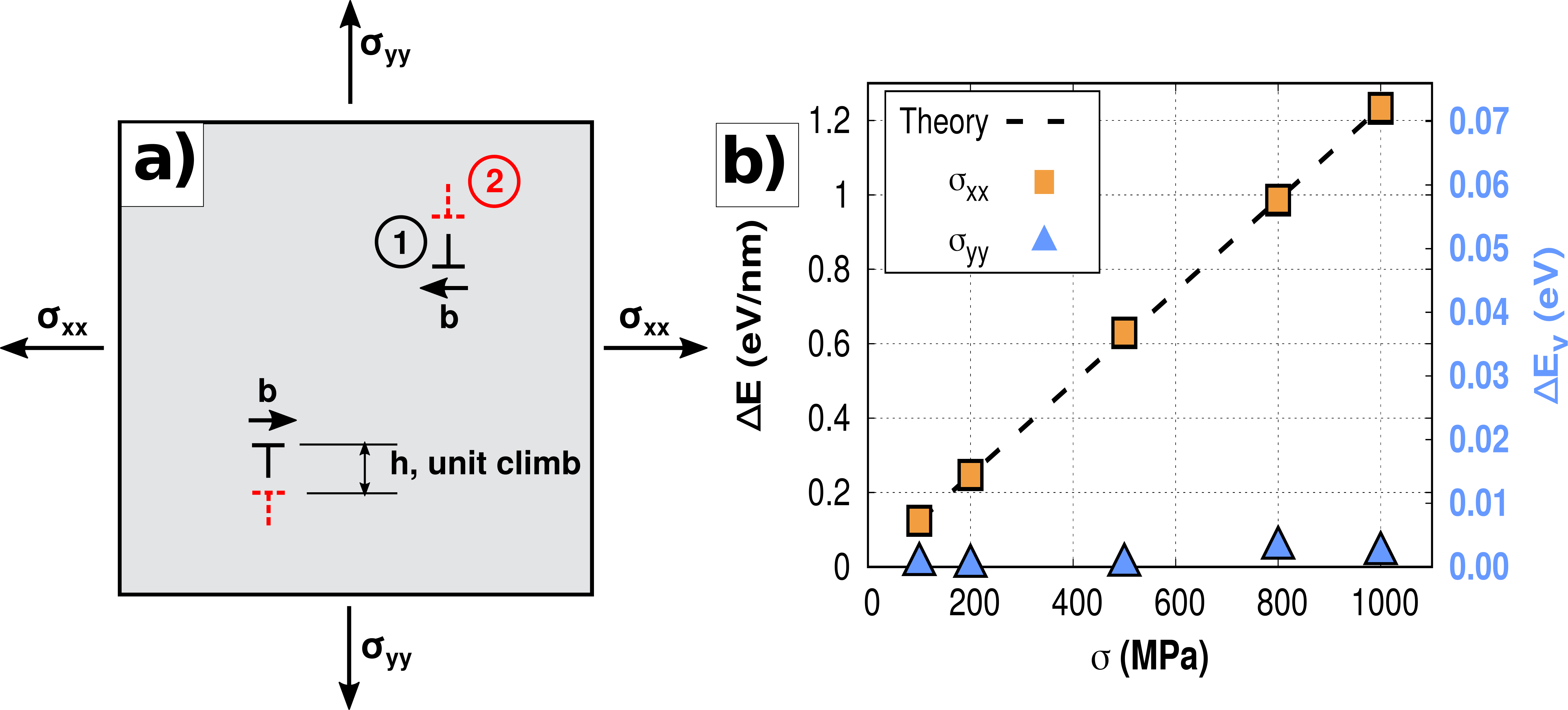}
\caption{\label{fig:climb} a) Sketch showing a unit climb process of an edge dislocation dipole with Burgers vector \textbf{b} from configurations 1 to 2. b) Energy change $\frac{\Delta E}{L}=-\frac{\delta W}{L}$ (left $y$ axis) and energy change per vacancy absorbed at the dislocation core (right $y$ axis) in the process of dislocation climb by a unit length $h=0.4$ nm.}
\end{figure}

Creep in metallic systems is the main plastic deformation mechanism at applied stresses below the yield strength and at intermediate to high homologous temperatures.\cite{HirthLothe, HullBacon} Understanding the basic processes involved in thermal creep is of paramount importance to design optimized materials for high temperature applications. Dislocations play a crucial role in the creep process; they act as sources and sinks of defects, which under stress leads to thermal creep resulting from dislocation climb. For a net deformation to exist, a bias between the absorption/emission propensity of defects among dislocations with different orientations relative to the applied stress needs to be present, i.e., dislocations oriented such that the external stress performs work as dislocation climbs will migrate at a faster rate compared to dislocations with a different orientation. Figure~\ref{fig:climb} sketches a unit climb mechanism of an edge dislocation dipole with Burgers vector \textbf{b}. 

Creep rate is determined by the net flux of defects emitted/absorbed from dislocation cores. Within linear response, fluxes are determined by the product of a mobility matrix (related to diffusion properties), and a force (related to gradients of chemical potentials). Both terms depend on the stress state of the system, although their relative sensitive to the action of the external stress in a given material is not known \textit{a priori}. The answer to this question for the particular case of edge dislocations in $\alpha$-Fe is the core of this paper.

The work done by the external stress per unit dislocation length, and the force on the dislocation are given by, \cite{HirthLothe}

\begin{equation}\label{eq:elas}
\frac{\delta W}{L}=-\sigma _{xx}bh, \quad \frac{F_{y_{el}}}{L}=\sigma _{xx}b
\end{equation}
%\begin{equation}
%\frac{F_{y_{el}}}{L}=\sigma _{xx}b
%\end{equation}
where $\sigma _{xx}$ is the $xx$ component of the applied stress and $h$ is the climbed height. $F_{y_{el}}$ is defined as $\delta W/h$, the change in energy as the dislocation moves a distance unit in the climb direction. In correspondence with the work done by the external stress, there is a change in the free energy of the material related to the emission/annihilation of vacancies at the dislocation core during climb. This change in free energy per unit dislocation length, and its associated osmotic force on the dislocation, are \cite{HirthLothe},
\begin{equation}
\frac{\delta G}{L}=\frac{\mu_vbh}{v_a}, \quad \textrm{with} \quad \mu_v= k_BT\ln \frac{c}{c^0}, %\quad \textrm{the vacancy chemical potential}
\end{equation}
\begin{equation}
\frac{F_{y_{os}}}{L}=-\frac{\mu_v b}{v_a}=-\frac{k_BTb}{v_a}\ln \frac{c}{c^0},
\end{equation}
where $\mu_v$ is the vacancy chemical potential, $v_a$ is the atomic volume, $c$ is the vacancy concentration, $c^0$ is a reference vacancy concentration which accounts for the interaction energy between the vacancy and the internal stresses\cite{HirthLothe}
\begin{equation}
  \label{interaction-energy}
c^0(\bm{r}) = \frac{1}{v_{a}}\exp{\left(-\frac{G^{\textrm{f}}-\mathbb{P}_{ij}\varepsilon_{ij}(\bm{r})}{k_{\textrm{B}} T}\right)}.
\end{equation}
In this equation vacancies are elastically described by their double force tensor $\bm{\mathbb{P}_{ij}}$,\cite{clouet_elastic_2018} with $\bm{\varepsilon}(\bm{r})$ the strain tensor at the location of the vacancy (summation over repeated indices is implied). The free energy of formation of the vacancy without any stress effect is denoted as $G^{\textrm{f}}$. Note that in thermal equilibrium the vacancy chemical potential is constant across the sample, while the vacancy concentration is not, as its formation energy is affected by the dislocation stress/strain field. %$F_{y_{os}}$ is again defined as $\delta G/h$, the change of energy for the dislocation to climb a unit length.

The climb force induced by the external stress, which favors vacancy absorption or emission, alters the concentration of vacancies in the vicinity of dislocations. This departure from equilibrium conditions creates a restoring force, the osmotic force, which tries to restore the concentration of vacancies at its equilibrium value.\cite{HirthLothe} Local equilibrium is obtained when these two forces are equal and opposite.
The total force on the dislocation at steady state is thus,

\begin{equation}
\frac{F_{y}}{L}=0=\frac{F_{y_{el}}}{L}+\frac{F_{y_{os}}}{L}=\sigma _{xx}b-\frac{k_BTb}{v_a}\ln \frac{c}{c^0},
\end{equation}
which results in the vacancy concentration close to the dislocation
\begin{equation}
  \label{eq-equilibrium-with-sigxx}
c(\bm{r}) = c^0(\bm{r})\exp \left( \frac{\sigma _{xx}v_a}{k_BT} \right).
\end{equation}

Note that, for the configuration presented in Fig.~\ref{fig:climb}(a), there is an elastic force only for non-zero $\sigma_{xx}$ regardless of the values of $\sigma_{yy}$. More generally, for a dislocation of Burgers vector $\bm{b}$ and line direction $\bm{\xi}$, the vacancy concentration is given by
\begin{equation}
  \label{eq-equilibrium-concentration-vacancies}
  c(\bm{r}) = c^0 (\bm{r}) \exp{\left(\frac{\left[(\bm{b} \cdot \bm{\sigma})\times\bm{\xi} \right] \cdot \left(\bm{b}\times \bm{\xi} \right)}{\left|\bm{b}\times \bm{\xi} \right|^2} \frac{v_a}{k_{\mathrm{B}} T}\right)}.
\end{equation}

%[\textcolor{blue}{The following paragraph could be removed if we needed some place; we should however keep Eint}]
Since the vacancy is isotropic at stable point,  its dipole tensor is simply $\mathbb{P}_{ij} = K \Delta V^{\mathrm{r}} \delta_{ij}$, with $K$ the bulk modulus, $\Delta V^{\mathrm{r}}$ the relaxation volume and $\delta_{ij}$ the Kronecker delta. The interaction energy becomes
\begin{equation}
  \label{eq-interaction-energy-isotropic}
  E^{\mathrm{int}}(\bm{r}) = -\mathbb{P}_{ij}\varepsilon_{ij}(\bm{r}) = (p^{\mathrm{a}}+p^{\mathrm{i}}(\bm{r})) \Delta V^{\mathrm{r}},
\end{equation}
where $p^{\mathrm{x}} = - \sigma_{ii}/3$ is the pressure and label $\textrm{x}$ refers either to the applied stress ($\textrm{x} = \textrm{a}$) or internal stress ($\textrm{x} = \textrm{i}$). $p^{\textrm{i}}$ is negligible sufficiently far from an isolated dislocation. Hence, Eq. (\ref{eq-equilibrium-with-sigxx}) becomes~\cite{HirthLothe} 
%[\textcolor{blue}{Eqs. 15.73-15.76 Hirth and Lothe}]
\begin{equation}
  \label{eq-equilibrium-ceq-hydrostatic-pressure}
  c = \frac{1}{v_a} \exp{\left(-\frac{G^{\textrm{f}} + p^{\mathrm{a}}\Delta V^{\mathrm{r}} - \sigma_{xx} v_a}{k_{\mathrm{B}}T}\right)}.
\end{equation}

In a general case, some components of the stress tensor will perform work while others will not. This asymmetry results in a bias for the fluxes of defects towards or from dislocations oriented dissimilarly. 

The first objective of this work is to reproduce this continuum result using a fully atomistic description, \textit{i. e.} that the change in internal energy of the sample produced by climb equals the work done by the external force. We use molecular statics (MS) and molecular dynamics (MD) simulations with an empirical interatomic potential for $\alpha$-Fe\cite{Ackland2004} as implemented in the LAMMPS code.\cite{LAMMPS} We create an edge dislocation dipole in a sample of 98.9$\times$90.5$\times$2.8 nm$^3$ and $2,147,700$ atoms with orientation $x=[111]$, $y=[1\bar{1}0]$, and $z=[11\bar{2}]$, with the Burgers vector in the $x$ direction and the dislocation tangent in the $z$ direction. The initial configuration contains two dislocations at exactly $[1/2L_x:1/4L_y]$ and $[1/2L_x:3/4L_y]$. The sample was annealed at 100 K and zero pressure for 25 ps and then minimized at zero pressure using a conjugate gradient algorithm with a $10^{-4}$ eV/\AA ~tolerance in atomic forces. Subsequently, the sample was relaxed to a minimum of enthalpy at different levels of independently applied normal stresses in the $x$ and $y$ directions, from 100 MPa to 1 GPa. The obtained configuration will be denoted as the initial sample. A second sample was generated with the dislocations in a configuration as if they would have climbed, i.e., at a position $[1/2L_x:1/4L_y-h]$ and $[1/2L_x:3/4L_y+h]$, with $h$ the unit climb distance (that was considered as 0.4 nm). The total number of atoms in this case is $2,147,652$. The relaxations to minimum enthalpy at different levels of normal stresses were repeated for this new configuration. The resulting structure is called final sample, and the energy will be reported as final energy. To calculate the work done by the external force, the final sample was brought back to the volume of the initial sample and minimized at constant volume. The difference between this last energy and the stored final energy is the work done by the external forces in the climb process. Figure~\ref{fig:climb}(b) shows the results obtained following this methodology compared to Eq.~\ref{eq:elas} (the sign of the energy was reversed to comply with the thermodynamic criterium of positive work if performed by the system). We observe that indeed, the atomistic results agree with the continuum approach, as they should if self-similarity of the dislocations performing climb holds. The figure also shows on the right $y$ axis the average energy change per absorbed vacancy at the dislocation core, which equals the vacancy chemical potential at the dislocation core ($\Delta E_v=\sigma _{xx}v_a$, with $v_a=11.64$ \AA$^3$ the atomic volume of $\alpha$-Fe)\cite{HirthLothe}. We note that the values of this last energy bias are small even at large levels of applied stress.  
%Note also that the bias is captured and that the system performs work if the dislocations are oriented with respect to the applied external stress to do so. Otherwise, as it is the case when $\sigma_{yy}$ is applied, the system does not perform any work.  
%\begin{figure}
%\includegraphics[width=0.8\columnwidth]{figures/EnergyBias.eps}
%\caption{\label{fig:bias} Energy change $\frac{\Delta E}{L}=-\frac{\delta W}{L}$ (left $y$ axis) and energy change per vacancy absorbed at the dislocation core (right $y$ axis) in the process of dislocation climb by a unit length $h=0.4$ nm. }
%
%\end{figure}

\begin{figure}
\includegraphics[width=0.7\columnwidth]{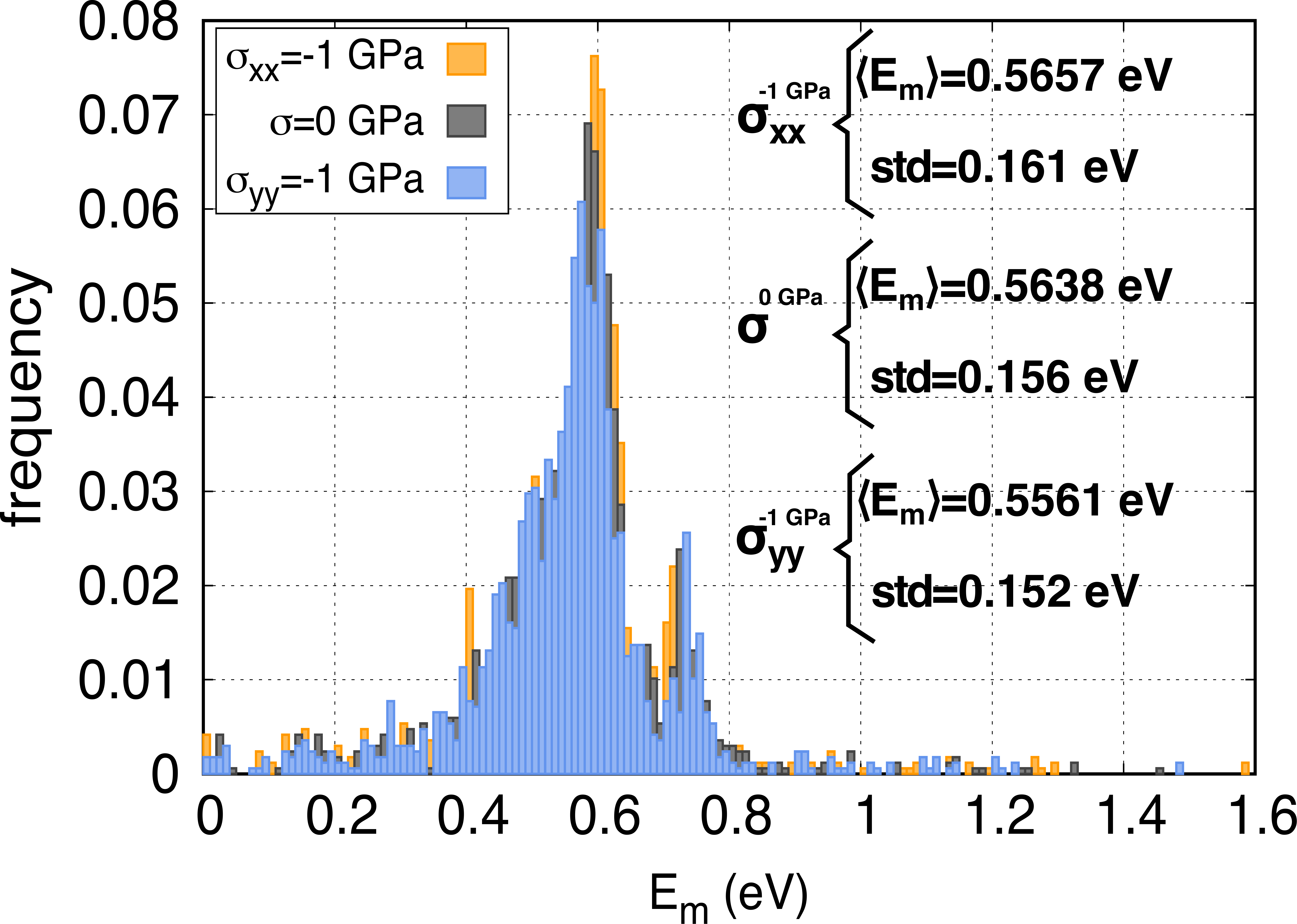}
\caption{\label{fig:neb}  Energy barrier distribution for vacancy migration around an edge dislocation core in $\alpha$-Fe depending on the applied stress.}
\end{figure}

Reaching the stationary state including defect emission is beyond the time scale accessible of MD. To circumvent this limitation we study the bias from a thermodynamic point of view, \textit{i. e.} by studying residence times of vacancies around the dislocation core using kinetic Monte Carlo (KMC) techniques.
With the atomistic description we calculated the energy barriers for a vacancy to migrate to all first nearest neighbor atoms at every site near the dislocation core. We used the nudged-elastic band (NEB) method \cite{Henkelman2000-2,Henkelman2000} in a cylindrical region of 1 nm in radius around the dislocation core. The calculations were performed at zero external stress, at $\sigma _{xx}=-1$ GPa and $\sigma _{yy}=-1$ GPa. The energy barrier distribution is shown in Fig.~\ref{fig:neb} where we readily notice that indeed the external stress modifies such distributions. The average and the standard deviation of the distributions are also given in the figure. We estimate that the maximum average barrier is obtained for a $\sigma _{xx}$ applied stress, i.e., when the dislocation is well-oriented to performed work, followed by the scenario with no external stress and last when the external stress is $\sigma _{yy}$. Moreover, the standard deviation follows a similar trend, i.e., the distribution is broader when the applied stress is the $\sigma _{xx}$, then when there is no applied stress and lastly when $\sigma _{yy}$ is applied. If one is to look just at the average energy barriers (for a similar pre-exponential factor), the impression would be that the vacancy will migrate slightly faster when the dislocation is not oriented to perform work, which at first glance might seem counterintuitive. 

To better understand how this distribution affects the vacancy transport, we used the resulting energy landscape in conjunction with a KMC algorithm to calculate defect mobilities. The rates were calculated following harmonic transition state theory ($\Gamma=\nu_0 \exp (-\Delta E/k_BT)$), where $\nu_0$ was taken as a constant pre-factor equal to $10^{13}$ s$^{-1}$, $\Delta E$ is the migration energy barrier (calculated for each jump previously with NEB), $k_B$ is the Boltzmann constant and $T$ the temperature. We performed thirty independent runs for each condition and calculated the average and standard deviation of the total time that the vacancy took to perform 10$^9$ jumps in each independent simulation. 
%This residence time is proportional to the defect concentration [\textcolor{blue}{Is it necessary to state this point? May be misleading, since here you are evaluating mobilities. After thinking more about that, I wonder if you don't want to suggest that you could in principle evaluate the effect of stress on the local concentration. Can you confirm this point?}]. 
The results for this dislocation configuration show that the total time spent by the defect to complete the simulations was similar in all three cases: $3.24018\cdot 10^{-5}\pm 1.45421\cdot 10^{-9}$ s, $3.24018\cdot 10^{-5}\pm 1.95703\cdot 10^{-9}$ s and $3.24020\cdot 10^{-5}\pm 1.97914\cdot 10^{-9}$ s for applied $\sigma _{xx}$, zero stress and $\sigma _{yy}$, respectively. Surprisingly, even though the external stress modifies the energy barrier distribution in these configurations, it does not significantly vary the defect diffusivity, and therefore, the residence time. 
%Interestingly, we see an anisotropic behavior in the site frequency, both in visits and time, with respect to the applied stress. A preference for the defect to reside on one side of the dislocation core versus the other develops for different boundary conditions (see Fig.~\ref{fig:barriers}). This effect does not correlate with a mere lattice effect \cite{Tome1982} since the atomic regions are broad enough to encompass several $\langle 111 \rangle$ directions. 

To test further these results, we have analyzed the vacancy diffusivity in the presence of a Volterra field of edge dislocations using a dipolar approximation as implemented in an Object KMC code (OKMC).\cite{vattre_non-random_2016,jourdan_continuous_2018} A model for point defect emission has been introduced in this code, which ensures proper local equilibrium near dislocations even when the effect of elastic interactions on the migration of point defects is taken into account. More details about this model, in particular its numerical implementation, are given in Ref.~\onlinecite{jourdan_modelling_nodate}. We briefly remind the derivation of the model for the sake of completeness.

The evolution equation for the concentration of vacancies near the dislocation can be written as 
\begin{equation}
  \label{eq-evolution-concentration}
  \frac{\mathrm{d}c(\bm{r})}{\mathrm{d}t} = \Gamma (\bm{r}) - \sum_{j = 1}^z \omega_{\cdot \rightarrow j} c(\bm{r}) + \sum_{j=1}^z \omega_{j \rightarrow \cdot} c(\bm{r}+\bm{h}_j) \eta_j,
\end{equation}
where $\omega_{\cdot  \rightarrow j}$ ($\omega_{j \rightarrow \cdot }$) is the jump frequency to (from) neighbour site $j$, located at $\bm{r}+\bm{h}_j$, $\eta_j = 1$ if $\bm{r}+\bm{h}_j$ is in the matrix and $\eta_j = 0$ if $\bm{r}+\bm{h}_j$ is in the capture region of the dislocation. $\Gamma$ is a source term to impose the local equilibrium concentration.

The jump frequency can be written as
\begin{equation}
  \label{eq-jump-frequency-vacancy}
  \omega_{\cdot \rightarrow j} = \nu_0 \exp{\left(-\frac{E^{\mathrm{m}}+E^{\mathrm{int,SD}}(\bm{r}+\bm{h}_j/2)-E^{\mathrm{int}}(\bm{r})}{k_{\mathrm{B}} T} \right)},
\end{equation}
where $E^{\mathrm{m}}$ is the migration energy in the bulk ($E^{\mathrm{m}} = 0.63~\mathrm{eV}$), without any effect of stress, and $E^{\mathrm{int,SD}}(\bm{r}+\bm{h}_j/2)$ and $E^{\mathrm{int}}(\bm{r})$ are the interaction energies with the applied and internal stresses at the saddle (between $\bm{r}$ and $\bm{r}+\bm{h}_j$) and equilibrium positions, respectively. At equilibrium ($\mathrm{d}c(\bm{r})/\mathrm{d}t = 0$), substituting $c$ (Eq.~(\ref{eq-equilibrium-concentration-vacancies})), assuming that the factor related to the Peach-Koehler force is the same for $\bm{r}$ and all positions $\bm{r}+\bm{h}_j$ and after some algebra we obtain
%
%
%
%
%, which is determined by writing Eq.~(\ref{eq-evolution-concentration}) at equilibrium:
%\begin{equation}
%  \label{eq-evolution-concentration-equilibrium}
%  G_v(\bm{r}) - \sum_{j=1}^z \omega_{\cdot \rightarrow j} c(\bm{r}) + \sum_{j=1}^z \omega_{j \rightarrow \cdot } c(\bm{r}+\bm{h}_j) \eta_j = 0,
%\end{equation}
%where $c$ is given by Eq.~(\ref{eq-equilibrium-concentration-vacancies}). This expression can be simplified if it is assumed that the factor related to the Peach-Koehler force is the same for $\bm{r}$ and all positions $\bm{r}+\bm{h}_j$:
%\begin{align}
%  \label{G-to-ensure-equilibrium}
%  G_v(\bm{r}) &= \left[\sum_{j=1}^z \omega_{\cdot \rightarrow j} c^0(\bm{r})  - \sum_{j=1}^z \omega_{j \rightarrow \cdot} c^0(\bm{r}+\bm{h}_j) \eta_j \right]  \times \nonumber \\
%  & \exp{\left(\frac{\left[(\bm{b} \cdot \bm{\sigma})\times\bm{\xi} \right] \cdot \left(\bm{b}\times \bm{\xi} \right)}{\left|\bm{b}\times \bm{\xi} \right|^2} \frac{v_a}{k_{\mathrm{B}} T}\right)}.
%\end{align}

%The jump frequency can be written as
%\begin{equation}
%  \label{eq-jump-frequency-vacancy}
%  \omega_{\cdot \rightarrow j} = \nu_0 \exp{\left(-\frac{E^{\mathrm{m}}+E^{\mathrm{int,SD}}(\bm{r}+\bm{h}_j/2)-E^{\mathrm{int}}(\bm{r})}{k_{\mathrm{B}} T} \right)},
%\end{equation}
%where $E^{\mathrm{m}}$ is the migration energy in the bulk ($E^{\mathrm{m}} = 0.63~\mathrm{eV}$), without any effect of stress and $E^{\mathrm{int,SD}}(\bm{r}+\bm{h}_j/2)$ is the energy at the saddle position between $\bm{r}$ and $\bm{r}+\bm{h}_j$.  Inserting Eq.~(\ref{eq-jump-frequency-vacancy}) in Eq.~(\ref{G-to-ensure-equilibrium})
%
%
\begin{multline}
  \label{eq-G-to-ensure-equilibrium-saddle}
   \Gamma (\bm{r}) = \frac{\nu_0 }{v_a}\exp{\left(-\frac{G^{\mathrm{f}}}{k_{\mathrm{B}}T}\right)} \exp{\left(-\frac{E^{\mathrm{m}}}{k_{\mathrm{B}}T}\right)} \times \\ \left[\sum_{j=1}^z \exp{\left(-\frac{E^{\mathrm{int,SD}}(\bm{r}+\bm{h}_j/2)}{k_{\mathrm{B}} T} \right)} (1-\eta_j)  \right] \times \\  \exp{\left(\frac{\left[(\bm{b} \cdot \bm{\sigma})\times\bm{\xi} \right] \cdot \left(\bm{b}\times \bm{\xi} \right)}{\left|\bm{b}\times \bm{\xi} \right|^2} \frac{v_a}{k_{\mathrm{B}} T}\right)}.
 \end{multline}

 It is worth noting that the emission rate following expression (\ref{eq-G-to-ensure-equilibrium-saddle}) at a given location is zero if there is no jump from this location to the capture region of the dislocation ($\eta_j = 1$ for all neighbors). The emission is therefore limited to a region around the dislocation, and hence spatial correlations are captured in the model. %For a dislocation whose capture region is a cylinder of radius $r_c$ ($r_c = 1~\textrm{nm}$), the emission region is bounded by a cylinder of radius $r_c+d_{\mathrm{max}}$, where $d_{\mathrm{max}}$ is the maximum jump distance.
 Furthermore, the emission rate does not depend on the interaction energy at stable sites but on the interaction energy at the saddle points.

The OKMC model discretizes a straight dislocation using cylindrical symmetry. \cite{jourdan_modelling_nodate} The emission rate inside an elementary volume $V_i$ is thus given by
\begin{equation}
  \label{eq-emission-rate}
  P_i = \Gamma (\bm{r}_i) V_i,
\end{equation}
The position $\bm{r}_i$ is chosen as the center of mass of the emission volume.

%The off-lattice OKMC approach requires a discretization of the emission region around the dislocation. For a straight dislocation, it has a natural cylindrical symmetry (Fig.~\ref{fig-schematics-emission-region-dislocation-segment}). The emission rate inside an elementary volume $V_i$ is given by
%\begin{equation}
%  \label{eq-emission-rate}
%  P_i = G_v(\bm{r}_i) V_i,
%\end{equation}
%where $i$ refers to the considered volume and $V_i = 1/2 (r_1+r_2)(r_2-r_1)\Delta \theta \Delta l$. The position $\bm{r}_i$ is chosen in the middle of the emission volume. More details about the model and its validation are given elsewhere \onlinecite{Thomas}
%
%\begin{figure}[htbp]
%  \centering
%  \includegraphics[width=0.6\columnwidth] {figures/cylinder_dislo-crop.pdf}
%  \caption{Schematics of the emission region around an edge dislocation segment. An elementary emission volume is shown in blue, the emission point is given by the blue dot.}
%  \label{fig-schematics-emission-region-dislocation-segment}
%\end{figure}

 To study the vacancy fluxes under stress, first the vacancy dipole tensors at the minimum and at the saddle point are obtained with the same interatomic potential used in the atomistic simulations ($\mathbb{P}_{\mathrm{SD}}$ corresponds to a jump in the $[111]$ direction):

\begin{align}
%$$
\mathbb{P}_{\mathrm{MIN}} &= \left (
\begin{matrix}
-2.80 & 0 & 0 \\
0 & -2.80 & 0 \\
0 & 0 & -2.80 \\
\end{matrix}
\right ) (\mathrm{eV}),
\nonumber \\
\mathbb{P}_{\mathrm{SD}} &=
\left (
\begin{matrix}
-4.60 & -1.81 & -1.81 \\
-1.81 & -4.60 & -1.81 \\
-1.81 & -1.81 & -4.60 \\
\end{matrix}
\right ) (\mathrm{eV}).
%$$
\end{align}

\begin{figure}
  \centering
  \includegraphics[width=0.7\columnwidth]{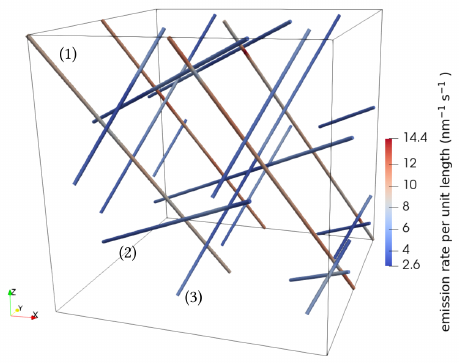}
  \caption{Emission rate (without absorption) of vacancies by dislocations at 773 K, per unit length of dislocation. The (200 nm)$^3$ box contains 3 straight dislocations, whose types are given in Table~\ref{tab-dislocation-types}.}
  \label{fig-emission-rate-dislocations}
\end{figure}

We have applied the OKMC model to a system at 773~K with three edge dislocations of different types (see Table ~\ref{tab-dislocation-types}) but of the same length in a $200\times 200\times 200$~nm$^3$ simulation box, whose edges coincide with the crystallographic axes $[100]$, $[010]$ and $[001]$ (Fig.~\ref{fig-emission-rate-dislocations}). In all cases, a uniaxial stress of magnitude $\sigma$ is applied along $\bm{t}_\sigma = 1/\sqrt{3}[111]$.

\begin{table*}[htbp]
  %\begin{ruledtabular}
    \begin{tabular}{c|c|c|c|c|c }
      dislocation type & $\bm{b}$ & $\bm{\xi}$ & $\displaystyle \frac{\left[(\bm{b} \cdot \bm{\sigma})\times\bm{\xi} \right] \cdot \left(\bm{b}\times \bm{\xi} \right)}{\left|\bm{b}\times \bm{\xi} \right|^2}$ & $\bm{\xi}\cdot \bm{t}_\sigma$ & $\beta$ (nm$^{-1}$s$^{-1}$)\\ \hline
      1                & $\bm{b} = \frac{a}{2}[111]$         & $\bm{\xi} = \frac{1}{6}[11\bar{2}]$ & $\sigma$ & 0 & $-0.10409\pm 0.00039 $\\
      2                & $\bm{b} = \frac{a}{2}[11\bar{1}]$   & $\bm{\xi} = \frac{1}{6}[2\bar{1}1]$ & $\frac{\sigma}{9}$ & $\frac{\sqrt{2}}{3}$ & $0.05110 \pm 0.00032$\\
      3                & $\bm{b} = \frac{a}{2}[11\bar{1}]$   & $\bm{\xi} = \frac{1}{6}[112]$ &  $\frac{\sigma}{9}$ & $\frac{2\sqrt{2}}{3}$ & $0.05299 \pm 0.00039$\\
    \end{tabular}
  %\end{ruledtabular}
  \caption{Dislocation types considered for the study of climb under stress and associated net absorption rate of vacancies $\beta$ (per unit length of dislocation, confidence interval given with a three-sigma rule on a set of 50 independent calculations).}
  \label{tab-dislocation-types}
\end{table*}

As a first step, we check that the local concentration close to each dislocation corresponds to the expected value. For this purpose, one single dislocation type is introduced in the box with zero initial vacancy concentration. Vacancies start to be emitted by the dislocation and their number increases, until a steady state is reached. The atomic fraction of vacancies in a box containing one of the three dislocations is $1.37\times 10^{-7} \pm 3\times 10^{-10}$, $7.34\times 10^{-8} \pm  1.7\times 10^{-10}$ and $7.32\times 10^{-8} \pm 4 \times 10^{-10}$, close to the theoretical values $1.33\times 10^{-7}$, $7.11\times 10^{-8}$ and $7.11\times 10^{-8}$ respectively, corresponding to Eq.~(\ref{eq-equilibrium-concentration-vacancies}) without internal stress field (a three-sigma criterion on a block-average method~\cite{Flyvbjerg1989} was used for the evaluation of the uncertainty). The small difference between the numerical and theoretical values may come from the discretized nature of the emission, although trends are correctly captured.

In a second type of simulation, vacancies are created in the bulk and diffuse in the matrix, until they are absorbed at one of the three dislocations. No vacancies can be emitted from dislocations. As shown in Refs.~\cite{Tome1982,Skinner1984,Woo1984}, the diffusion of vacancies becomes anisotropic due to the applied stress. The absorption rate of vacancies by dislocations depends on the orientation of the dislocation line with respect to the direction of the uniaxial stress $\bm{t}_\sigma$. The properties of the dipole tensor of vacancies at saddle point are such that a dislocation whose line direction is along $\bm{t}_\sigma$ captures more vacancies that a dislocation orthogonal to $\bm{t}_\sigma$. This prediction is confirmed by our numerical calculations: the three dislocations, which have different values of $\bm{\xi}\cdot \bm{t}_\sigma$ (Tab.~\ref{tab-dislocation-types}), do not absorb the same quantity of vacancies. Dislocation (3) absorbs 10\% more vacancies than dislocation (2) and 12\% more vacancies than dislocation (1), in line with the theoretical predictions. Therefore, anisotropic diffusion plays a role in certain configurations, difficult to probe with atomistic simulations. Note that these results do not depend on the incoming vacancy flux.

Finally, a last type of simulation is performed with the three dislocation types in the sample and vacancies generated thermally at the dislocation lines. A comparison of emission rates at dislocations (Fig.~\ref{fig-emission-rate-dislocations}) shows that dislocation (1) imposes a large local vacancy concentration, which leads to a net flux of vacancies to dislocations (2) and (3). Dislocation (3) emits more vacancies than dislocation (2) to sustain the same local concentration. This higher emission rates compensates for the higher absorption rate due to stress-induced anisotropic diffusion. The net absorption rates per unit length of dislocation are given in Table~\ref{tab-dislocation-types}. They are shown to be different for dislocations (2) and (3) due to the stress-induced anisotropic diffusion. The effect is, however, very small compared to the thermodynamic driving force which is responsible for the large difference between dislocation (1) and dislocations (2) and (3).

These results show that the main source of bias in $\alpha$-Fe as described by the interatomic potential used in this study comes from the probability of defect absorption or emission such that work is performed. \cite{HirthLothe} Anisotropic diffusion induced by local stresses \cite{lau_atomistic_2009,kabir_predicting_2010,baker_multiscale_2016}, lattice effects \cite{Tome1982,Skinner1984,Woo1984} or second order coupling between the external and dislocation fields \cite{Bullough1975} play a lower order role.

As we have seen, the absorption/emission bias is in the order of tens of meV (see Fig.~\ref{fig:climb}(b)). This result makes daunting the direct use of MD or even advanced methods such as accelerated MD \cite{Voter1997,Voter1998,TAD} or adaptive kinetic Monte Carlo algorithms \cite{kART2008,Xu2011,kabir_predicting_2010} to study the climb process, since capturing the bias with these methodologies beyond statistical noise will imply a significant computational effort.

In summary, we have validated the thermodynamic driving force for dislocation climb derived from elasticity theory. Using MS/MD simulations we have found the bias for defects to be absorbed/emitted from a dislocation depending on its relative orientation with respect to the applied stress. We have computed the migration energy barriers for a vacancy to jump inside a cylinder of 1 nm in radius around the dislocation core, for different external conditions: at zero stress, $\sigma _{xx}=$ 1 GPa and $\sigma _{yy}=$ 1 GPa. The distribution of barriers show that indeed there is an effect of the external stresses on the average and standard deviation of the energy barrier distributions. The three sets of barriers were used in conjunction with a KMC algorithm to study the mobility of the vacancy. Surprisingly, the mobilities obtained through this methodology result in similar values of residence times (related to the vacancy concentration), indistinguishable in a statistical sense. The energy bias obtained for the dislocation climb process is of the order of tens of meV, which makes daunting its dynamical calculation with available atomistic methods. Therefore we have verified the results with an OKMC model using a first order approximation to the vacancy interaction energies in the presence of the strain field generated by several dislocations. Both emission and absorption were considered. The conclusion is the same as for the atomistic approach: the main effect is the thermodynamic driving force for climb with a much lower effect of anisotropic diffusion induced by local stresses. 

E.M., A.C. and A.A want to thank John P. Hirth for useful discussions. E.M. acknowledges enlightning discussions with Profs. William Curtin and Jaime Marian. E.M. thanks support from the startup package at Clemson University. Clemson University is acknowledged for generous allotment of compute time on Palmetto cluster. T.J. acknowledges that his work has been partly carried out within the framework of the EUROfusion Consortium and has received funding from the Euratom research and training programme 2014-2018 under grant agreement no. 633053. The views and opinions expressed herein do not necessarily reflect those of the European Commission.
%

%\bibliographystyle{elsarticle-num}
%\bibliography{totalbiblioNEW}

\end{document}